\documentclass[12pt]{iopart}
\usepackage{graphicx}
\usepackage{setstack,cite}
\begin{document}


\title[Linearization of isochronous centre]{On the linearization of isochronous centre of a modified Emden equation with linear external forcing}
\author{R~Mohanasubha, M~I~Sabiya Shakila and M~Senthilvelan}%
\address{Centre for Nonlinear Dynamics, School of Physics,
Bharathidasan University, Tiruchirappalli - 620 024, India }







\begin{abstract}
In this work, we carry out a detailed study on the linearization of isochronous centre of a modified Emden equation with linear external forcing. We construct inverse integrating factor and time independent first integral for this system through Darboux method. To linearize the isochronous centre we explore a transverse commuting dynamical system and its first integral. With the help of first integrals of the original dynamical system and its transverse commuting system we derive the linearizing transformation and reduce the nonlinear system into linear isochronous one. We also point out certain mathematical structures associated with this dynamical system.
\end{abstract}
\hspace{2.5cm}Keywords: Darboux polynomials, Linearization, Isochronous centre.

\section{Introduction}
The aim of this paper is to study the linearization of isochronous centre of a modified Emden equation with linear external forcing, namely~\cite{first,nuci1,nuci2,partha,rana}
\begin{equation}
\ddot{x}+(2m+3)x^{2m+1}\dot{x}+x+x^{4m+3}=0,\label{j1}
\end{equation}
where $m$ is a non-negative integer and overdot denotes differentiation with respect to $t$. For the present study,  we consider the above equation in the following equivalent form, that is~\cite{iacono}
\begin{equation}
\dot{x}=y-x^{2m+2}=p(x,y),~~~\dot{y}=-x-x^{2m+1}y=q(x,y).\label{l1}
\end{equation}
The dynamical vector field of (\ref{l1}) is given by
\begin{equation}
X=(y-x^{2m+2})\frac{\partial} {\partial{x}}-(x+x^{2m+1}y)\frac{\partial} {\partial{y}}.\label{p1}
\end{equation}
Eq.(\ref{l1}) describes isochronous oscillations, that is the frequency of oscillations of (\ref{l1}) is independent of the amplitude~\cite{iacono,cha1}. The origin $0$ of system (\ref{l1}) is a centre if all the orbits in a neighborhood are closed. A centre is called an isochronous centre if the period of oscillations is same for all these orbits~\cite{isocen}. To show the period of oscillations is same in this system, let us consider the case $m=0$ in (\ref{l1}). The resultant equation of motion reads
 \begin{equation}
\dot{x}=y-x^2,~~~\dot{y}=-(x+xy).\label{z1}
\end{equation}
Now introducing the polar coordinates, $x=r\cos \theta$ and $y=r\sin \theta$, equation (\ref{z1}) can be brought to the form
\begin{equation}
\dot{r}=-r^2\cos\theta,~~~\dot{\theta}=-1 \label{g1}
\end{equation}
which is symmetric under $(r,~\theta)\rightarrow(-r,~\theta+\pi)$. Integrating the second equation in (\ref{g1}) we find the period of oscillations of the system is $T=2\pi$. In fact, the period of oscillations of (\ref{l1}) turns out to be the same constant, $2\pi$, for all positive integer values of $m$ which can be confirmed by straightforward calculations. In other words the dynamical system (\ref{l1}) has an isochronous centre at the origin~\cite{polar,po}. We also confirm the periodicity of the system by rewriting (\ref{z1}) in the complex form, that is $\dot{z}=-i(z+\frac{(z^2+|z|^2)} {2i})$, where $z=x+iy$, and calculating the period of trajectory $T$. Doing so, we find
\begin{equation}
\displaystyle T=\int \frac{dz} {i(-z-\frac{(z^2+|z|^2)} {2i})}=2\pi Res\bigg(\frac{1} {-z-\frac{(z^2+|z|^2)} {2i}},~0\bigg)= 2\pi.
\end{equation} Here also we observe that for all positive integer values of $m$ the origin is an isochronous centre~\cite{polar,po}.\\

 Now we investigate the linearization of isochronous centre of (\ref{l1}). The linearizability can be achieved by constructing a transverse commuting vector field $Y$ for $X$~\cite{toni,garcia,guha,f1,gine}. The necessary linearizing transformation is then obtained from the first integral of the dynamical system (\ref{l1}) and its transverse commuting dynamical system with the aid of the theorem proposed by Garcia and Maza ~\cite{garcia}. For the sake of completeness we also derive the general solution of the nonlinear ODE (\ref{l1}) from the solution of the linear ODEs.

 We intend to undertake this study due to the interest shown on this model (\ref{j1}) over the past few years in various contexts. In the following, we recall briefly the research activities revolving around the model (\ref{l1}). Chandrasekar et al have identified Eq.(\ref{j1}) as an integrable equation while applying Prelle-Singer procedure to the general class of Li$\mathrm {\acute{e}}$nard equation~\cite{first}. In a consequent study Eq.(\ref{j1}) with $m=0$ has been shown to admit $(1)$ a conservative Hamiltonian description with amplitude independent frequency of oscillations~\cite{cha1}. Recently Iacono and Russo have shown that the model (\ref{l1}) admits isochronous orbits around the origin and derived first integral, explicit general solution~\cite{iacono}. The second order nonlinear ODE can be transformed to simple harmonic oscillator equation through a nonlocal transformation~\cite{cha1,cha2}. The Hamiltonian dynamics of this case was also discussed in detail in Ref~\cite{cha2}. The method of integrating the second order ODE (\ref{j1}) was discussed by several authors, see for example Refs.~\cite{anna2,anna3,blum,car}. Apart from the classical studies, very recently the quantization of the  $m=0$ case has also been considered by Chithiika Ruby et al and the result demonstrates that the energy spectrum of this model is same as that of the harmonic oscillator~\cite{ruby}. 
Interestingly the $N$-dimensional integrable generalization of this equation was also studied by Gladwin Pradeep et al~\cite{glad}. The necessary and sufficient condition for the isochronicity of the Li$\mathrm{\acute{e}}$nard type system 
\begin{equation}
\dot{x}=y-xB(x),\;\;\dot{y}=-C(x)-yB(x),
\end{equation}
with $x^2B(x)=\int xf(x)dx$ and $C(x)=g(x)-x^2B^2(x)$ is also discussed in Ref~\cite{sabali}.

The plan of the paper is as follows. In the following section, we describe the method of finding inverse integrating factor and first integral for the dynamical system (\ref{l1}). We also discuss the construction of transverse commuting dynamical vector field. In sec. 3, we linearize the nonlinear model (\ref{l1}) into a linear isochronous one. We enumerate the results for three cases (i)~$m=0$,~(ii)~$m=1$ and (iii)~$m=arbitrary$, in this section. Finally, we present our conclusions in sec. 4.
\section{\bf Methodology}
\label{sec:1}
Let us consider the planar polynomial differential system (\ref{l1}) and its dynamical vector field $X$ given in Eq. (\ref{p1}).
We recall here that the main methods to study isochronous centre are classified in two categories. The first one says that a centre of the analytic system (\ref{l1}) is said to be isochronous if and only if there exists a commuting vector field $Y$ of the form~\cite{gine,saba,new1}
\begin{equation}
Y=r(x,y)\frac{\partial} {\partial{x}}+s(x,y)\frac{\partial} {\partial{y}},\label{ycom}
\end{equation}
such that $[X,Y]=XY-YX=0$. The commutation between (\ref{p1}) and (\ref{ycom}) yields
\begin{eqnarray}\hspace{-1.0cm}
(y-x^{2m+2})r_x-(x+x^{2m+1}y)r_y+(2m+2)x^{2m+1}-s&=&0,\label{new1}\\\hspace{-1.0cm}
(y-x^{2m+2})s_x-(x+x^{2m+1}y)s_y+r(1+(2m+1)yx^{2m})+sx^{2m+1}&=&0.\label{new2}
\end{eqnarray}
Solving Eqs.(\ref{new1}) and (\ref{new2}) one can obtain the explicit form of $r$ and $s$ from which one can fix the transverse dynamical system
\begin{equation}
\dot{x}\;=\;r(x,y),\;\;\;\;\dot{y}\;=\;s(x,y).\label{e3}
\end{equation}
The second one says that a centre of an analytic system is isochronous if and only if there exists a near-identity analytic change of coordinates $(u,v)=(x+o(|(x,y)|), y+o(|(x,y)|))$ transforming the system (\ref{l1}) into a linear isochronous system, that is
\begin{equation}
\dot{u}=v,~~~\dot{v}=-u.\label{har1}
\end{equation}
However, finding linearizing transformation for the given equation is often cumbersome. Recently, Garcia and Maza have given a procedure to derive the linearizing transformation. In this connection, they have proved the following theorem \cite{garcia}:\\
{\bf Theorem} Let $X=p\frac{\partial} {\partial{x}}+q\frac{\partial} {\partial{y}}=(-y+...)\partial_x+(x+...)\partial_y$ and $Y=r\frac{\partial} {\partial{x}}+s\frac{\partial} {\partial{y}}=(x+...)\partial_x+(y+...)\partial_y $ be two analytic vector fields in a neighborhood $U$ of the origin such that $[X,Y]=0$. Then, a near-identity change of variables $u=x+...,v=y+...$, analytic in $U$ that linearizes $X$ is obtained from:
\begin{equation}
u=\frac{\sqrt{f(H)}g(I)} {\sqrt{1+g^2(I)}},\;\;\;\;\;v=\frac{\sqrt{f(H)}} {\sqrt{1+g^2(I)}},\label{ee1}
\end{equation}
where $f$ and $g$ are two functions such that $f(H(x,y))=x^2+y^2+...$ and $g(I(x,y))=(y+...)/(x+...)$. 

Here $H$ and $I$ are time independent first integrals associated with the inverse integrating factor $X\wedge Y$ of the dynamical system (\ref{l1}) and its transverse commuting dynamical system (\ref{e3}) respectively. $X\wedge Y$ represents the wedge product which is given by $ps-qr$.

The question now boils down in finding (i) the time independent first integral of the dynamical system (\ref{l1}) and (ii) a transverse commuting vector field $Y$ and its first integral. As far as the first point is concerned even though there exists several methods in the literature to identify time independent first integral of the given dynamical system~\cite{anna2,bluman,sir} here we adopt the Darboux method which is best suited for the present problem. In the following, we describe this method very briefly, for more details one may refer Refs.~\cite{toni,dar1,ffl,fin}.

In Darboux's method, the integrating factor associated with the given planar equation $\dot{x}\;=\;p(x,y)$ and $\dot{y}\;=\;q(x,y)$,
\begin{equation}
R=\displaystyle\prod_if_{i}^{n_i},\label{if}
\end{equation}
where $f_i$'s are Darboux polynomials, can be obtained by solving the first order partial differential equation
\begin{equation}
p(x,y)\frac{\partial{f}} {\partial{x}}+q(x,y)\frac{\partial{f}} {\partial{y}}=g(x,y)f,
\end{equation}
where $g(x,y)$ is the cofactor of the Darboux polynomials. In the above $n'_i$s are rational numbers which can be calculated from
\begin{equation}
\sum_in_ig_i=-(p_x+q_y).\label{r1v}
\end{equation}
The inverse integrating factor $V$ is then determined from the expression
\begin{equation}
V=\frac{1} {R}=f_1^{n_1}f_2^{n_2}.\label{iif}
\end{equation}
From this inverse integrating factor one can proceed to construct the time independent first integral of the given planar system by evaluating the integral~\cite{garcia}

\begin{equation}
H=\int \frac{-qdx+pdy} {V}. \label{eq3}
\end{equation}

The next task is to determine the transverse commuting dynamical system (\ref{e3}). To determine the functions $r$ and $s$ we recall the identity~\cite{chav}
\begin{equation}
V=ps-qr\label{inv}.
\end{equation}
Since $p,~q$ and $V$ are known at this stage, we assume a specific ansatz for the functions $s$ and $r$ and substitute this ansatz in (\ref{inv}) and express the function $s$ in terms of $r$. We then insert these expressions either in (\ref{new1}) or (\ref{new2}) and determine the explicit form of $r$ and $s$. Finally, we confirm the obtained functions $r$ and $s$ also satisfy the second commuting requirement condition, that is (\ref{new2}) or (\ref{new1}). Since $V$ also forms the inverse integrating factor for the transverse commuting system~\cite{chav}, the first integral of the later can be derived by evaluating the integral
\begin{equation}
I=\int \frac{-sdx+rdy} {V}.\label{eq2}
\end{equation}

Finally, by appropriately choosing the functions $f(H)$ and $g(I)$ one can obtain the necessary linearizing transformation from (\ref{ee1}). In the new variables the original nonlinear ODE becomes the linear harmonic oscillator equation.


\section{\bf MEE equation}

In this section, we present the method of linearizing the isochronous centre of (\ref{l1}). To make our studies systematic, we first consider the case $m=0$ in (\ref{l1}) and present the method of finding transverse commuting vector field, first integral and the necessary linearizing transformations. We then consider the case $m=1$ and consolidate the results. Finally, we generate the results to the case $m=arbitrary$.
\subsection{\bf Case m=0}
To begin with let us consider the case $m=0$ in (\ref{l1}) (vide Eq.(\ref{z1})).
The associated dynamical vector field $X$ takes the form
\begin{equation}
X=(y-x^2)\frac{\partial} {\partial{x}}-x(1+y)\frac{\partial} {\partial{y}}.
\end{equation}
The Darboux polynomials of (\ref{z1}) can be determined by solving the partial differential equation
\begin{equation}
(y-x^2)\frac{\partial{f}} {\partial{x}}-(x+xy)\frac{\partial{f}} {\partial{y}}=g(x,y)f.\label{v1}
\end{equation}
To solve equation (\ref{v1}) we assume $f$ is a quadratic polynomial in $y$ with coefficients are unknown function of $x$, that is $f=f_0(x)+f_1(x)y+f_2(x)y^2$, where $f_0, f_1$ and $f_2$ are to be determined and $g$ is of the form $g=g_1x$ with $g_1$ is a constant. Substituting this ansatz in (\ref{v1}) and solving the resultant equation we find the necessary Darboux polynomials are of the form $(i)~f=x^2+y^2$ and $(ii)~f=-2-2y$ with co-factors $(i)~g=-2x$ and $(ii)~g=-x$ respectively. Substituting these expressions in $\sum_in_ig_i=-(p_x+q_y)$, we find $n_1=-1$ and $n_2=-1$. Since the Darboux polynomials and the exponents are now known the integrating factor can be fixed from (\ref{iif}), which in turn reads
\begin{equation}
R=-(2+2y)^{-1}(x^2+y^2)^{-1}.
\end{equation}
So that the inverse integrating factor takes the form
\begin{equation}
V=-2(1+y)(x^2+y^2).\label{iif0}
\end{equation}
Using this inverse integrating factor we can determine the first integral associated with the dynamical system (\ref{z1}) by evaluating the integral given in (\ref{eq3}). We find the resultant first integral, $H$, is of the form
\begin{equation}
H=\frac{(1+y)^2} {(x^2+y^2)}.\label{nb2}
\end{equation}
One can check unambiguously that the time derivative of $H$ is equal to zero, that is $\frac{dH} {dt}=0$.

Now we determine the transverse commuting vector field from the relations $V=ps-qr$ and (\ref{new1})-(\ref{new2}). To obtain $s$ and $r$, we make an ansatz for these two functions in the form  $s=s_0(x)+s_1(x)y+s_2(x)y^2$ and $r=r_0(x)+r_1(x)y+r_2(x)y^2$, where $r_i$'s and $s_i$'s, $i=0,1,2,$ which are all to be determined. Substituting these forms in the expression $V=ps-qr$ we get
\begin{equation}
-2(1+y)(x^2+y^2)=(y-x^2)(s_0+s_{1}y+s_{2}y^2)+(x+xy)(r_0+r_{1}y+r_{2}y^2).
\end{equation}
Equating the coefficients of different powers of $y$ on both sides, we find
\begin{eqnarray}
s_0(x)&=&-r_0x-r_1x-4x^2-r_1x^3-r_2x^3-2x^4-r_2x^5,\nonumber\\
s_1(x)&=&-2-r_1x-r_2x-2x^2-r_2x^3, \nonumber\\
s_2(x)&=&-2-r_2x.\label{ss1}
\end{eqnarray}
In other words, we have expressed the coefficient functions $s_i$'s in terms of $r_i$'s and reduced the unknowns.
Now substituting (\ref{ss1}) in (\ref{new1}) (with $m=0$) we get a set of differential equations for the functions $r_i$'s, $i=0,1,2$. Solving them consistently and choosing some of the integration constants are zero, for the sake of convenience, we arrive at the following form for $s$ and $r$, namely
\begin{eqnarray}
s=-2(y+y^2),\;\;\; r=-2(x+xy).\label{b1}
\end{eqnarray}

These particular solutions also satisfy the equation (\ref{new2}). Thus the commuting transversal system to the equation (\ref{z1}) is found to be
\begin{eqnarray}
\dot{x}=-2(x+xy),\;\;\; \dot{y}=-2(y+y^2).\label{ceq1}
\end{eqnarray}
The associated transversal commuting vector field reads $Y=-2(x+xy)\frac{\partial} {\partial{x}}-2(y+y^2)\frac{\partial} {\partial{y}}$.\\
Since (\ref{iif0}) also forms an inverse integrating factor for the commuting system the first integral of the later can be constructed by substituting the expressions (\ref{iif0}) and (\ref{b1}) in (\ref{eq2}) and evaluating the resultant integral. The resultant integral reads
\begin{equation}
I=tan^{-1}\bigg(\frac{y} {x}\bigg)-tan^{-1}\bigg(\frac{x} {y}\bigg).\label{nb1}
\end{equation}

From the integrals $H$ and $I$ (vide Eq.(\ref{nb2}) and (\ref{nb1})), we can proceed to determine the linearizing transformation. To do so we consider
\begin{equation}
f(H)=\frac{(1+y)^2} {x^2+y^2},\;\;g(I)=\frac{y} {x} \label{tr}.
\end{equation}
Substituting (\ref{tr}) in (\ref{ee1}) we find the linearizing transformation is of the from
\begin{equation}
u=\frac{y(1+y)} {(x^2+y^2)},\;\;\;v=\frac{x(1+y)} {x^2+y^2}.\label{tra}
\end{equation}
In terms of these new variables, $u$ and $v$, the dynamical system (\ref{z1}) reads
\begin{equation}
\dot{u}=v,\;\;\dot{v}=-u.\label{ha1}
\end{equation}

From the solution of the linear system (\ref{ha1}) we can construct the solution of the nonlinear system (\ref{z1}). Doing so we find
\begin{eqnarray}
x(t)=\frac{\sin(t-t_0)} {C-\cos(t-t_0)},~~~ y(t)=\frac{\cos(t-t_0)} {C-\cos(t-t_0)},
\end{eqnarray}
where $C$ and $t_0$ are integration constants.

\subsection{\bf Case m=1}
Now we consider the case $m=1$ in (\ref{l1}). The underlying dynamical system and its associated vector field are given by
\begin{eqnarray}
\dot{x}=y-x^4,~~\dot{y}=-x-x^3y\label{m=1}
\end{eqnarray}
and
\begin{equation}
X=(y-x^4)\frac{\partial} {\partial{x}}-(x+x^3y)\frac{\partial} {\partial{y}}.\label{ne1}
\end{equation}
To determine the Darboux polynomials $(f)$ and cofactors of (\ref{m=1}) we assume $f$ is a cubic polynomial in $y$, that is $f=f_0(x)+f_1(x)y+f_2(x)y^2+f_3(x)y^3$, where the coefficient functions are to be determined and $g$ is of the form $g_1x^3$ with $g_1$ is a constant. By solving the equation $(y-x^4)\frac{\partial{f}} {\partial{x}}-(x+x^3y)\frac{\partial{f}} {\partial{y}}=g_1x^3f$, we find the necessary Darboux polynomials are in the form
\begin{equation}
(i)~f=x^2+y^2,\;\;\;(ii)~f=-6(1+3x^2y+2y^3)
\end{equation}
with
\begin{equation}
(i)~g=-2x^3,\;\;\;(ii)~g=-3x^3.
\end{equation}
With these expressions, Eq.(\ref{r1v}) is evaluated to yield $n_1=n_2=-1$. The inverse integrating factor of the dynamical system (\ref{m=1}) is found to be
\begin{equation}
V=-6(x^2+y^2)(1+3x^2y+2y^3)\label{iif1}.
\end{equation}
The first integral $H$ of (\ref{m=1}) can be constructed by substituting the expressions $p=(y-x^4),\;q=-(x+x^3y)$ and the function $V$ given in (\ref{iif1}) and evaluating the resultant integral (\ref{eq3}). For the present case we find
\begin{equation}
H=\frac{(1+3yx^2+2y^3)^2} {(x^2+y^2)^3}\label{fi1}.
\end{equation}

Now we proceed to determine the transverse commuting vector field to (\ref{ne1}). As we did in the previous case, from Eqs.(\ref{inv}) and (\ref{new1}) we try to fix the explicit form of $r$ and $s$. Here also we assume the unknown functions $r$ and $s$ are polynomials in $y$ with unknown functions in $x$, that is $r=r_0(x)+r_1(x)y+r_2(x)y^2+r_3(x)y^3+r_4(x)y^4$ and $s=s_0(x)+s_1(x)y+s_2(x)y^2+s_3(x)y^3+s_4(x)y^4$. As we did earlier, we first express the functions, $s'_i$s, $i=0,1,2,3,4$, in terms of the coefficients $r'_i$s using (\ref{inv}). We then substitute the obtained forms of $s'_i$s in Eq.(\ref{new1}) and obtain the explicit form of $s$ and $r$. After a lengthy calculation we arrive at
\begin{eqnarray}
r=-6x(1+3x^2y+2y^3),~ s=-6y(1+3x^2y+2y^3).
\end{eqnarray}
These expressions also satisfy (\ref{new2}).
Thus the transverse commuting dynamical system is found to be
\begin{eqnarray}
\dot{x}=-6x(1+3x^2y+2y^3),~ \dot{y}=-6y(1+3x^2y+2y^3).\label{ceq2}
\end{eqnarray}
Since $V$ (vide Eq.(\ref{iif1})) is also an inverse integrating factor for this transverse dynamical system, (\ref{ceq2}), the first integral of the later can be obtained from (\ref{eq2}) itself. Interestingly, we find that the time independent first integral of (\ref{ceq2}) is nothing but the one which we found earlier in $m=0$ case, that is Eq.(\ref{nb1}).

Now fixing $f(H)=H$ and $g(I)=\frac{y} {x}$, where the first integrals $H$ and $I$ are given in equations (\ref{fi1}) and (\ref{nb1}) respectively, we find the necessary linearizing transformation is of the form
\begin{equation}
u=\frac{y(1+3x^2y+2y^3)} {(x^2+y^2)^2},\;\;\;\;\;v=\frac{x(1+3x^2y+2y^3)} {(x^2+y^2)^2}.\label{tra2}
\end{equation}
Under this transformation (\ref{tra2}), Eq.(\ref{m=1}) becomes linear harmonic oscillator equation (\ref{har1}). From the solution of the latter we find the following form of oscillatory solutions to the equation (\ref{m=1}), that is
\begin{eqnarray}
x(t)&=&\frac{\sin(t-t_0)} {(C-\cos (t-t_0)(2+\sin^2(t-t_0)))^{1/3}},\nonumber \\
y(t)&=&\frac{\cos(t-t_0)} {(C-\cos (t-t_0)(2+\sin^2(t-t_0)))^{1/3}},
\end{eqnarray}
where $C$ and $t_0$ are integration constants.

\subsection{\bf Case m=arbitrary}
Now we consider the case $m=arbitrary$.
The Darboux polynomials of (\ref{l1}) can be determined by solving the following partial differential equation, namely
\begin{equation}
(y-x^{2m+2})\frac{\partial f} {\partial{x}}-(x+x^{2m+1}y)\frac{\partial f} {\partial{y}}=g(x,y)f.
\end{equation}
By induction we find that the Darboux polynomials are of the form
\begin{equation}
(i)~f=(x^2+y^2),\;\;\;(ii)~f=-(4m+2)[1+\sum_{r=0}^m A_{mr}yx^{2r}(x^2+y^2)^{m-r}]
\end{equation}
with the cofactors $g$ as $(i)~g=-2x^{(2m+1)} $ and $(ii)~g=-(2m+1)x^{(2m+1)}$. Here $A_{mr}$ is a constant which can be written in the following compact form
\begin{equation}
A_{mr}\equiv\frac{2^{2(m-r)}(m!)^2(2r)!} {(2m)!(r!)^2}.\label{amr}
\end{equation}
 Proceeding further we find that the integers $n_1$ and $n_2$ which satisfy the condition $\sum_in_ig_i=-(p_x+q_y)$ are nothing but $n_1=n_2=-1$, as in the subcases. The inverse integrating factor $V$ associated with these Darboux polynomials is given by
\begin{equation}
V=-(4m+2)(x^2+y^2)[1+\sum_{r=0}^m A_{mr}yx^{2r}(x^2+y^2)^{m-r}].\label{k1}
\end{equation}
Now evaluating the integral (\ref{eq3}) with $p=y-x^{2m+2}$, $q=-x-x^{2m+1}y$ and $V$ as given in (\ref{k1}) we find that the first integral of (\ref{l1}) is of the form
\begin{equation}
H=\frac{[1+\sum_{r=0}^n A_{mr}yx^{2r}(x^2+y^2)^{m-r}]^2} {(x^2+y^2)^{2m+1}}.
\end{equation}

The remaining task is to determine a commuting transverse vector field $Y$ to the dynamical vector field $X$. To explore it we follow the same procedure described in the sub-cases, $m=0$ and $m=1$ respectively, with suitable changes in the ansatz for the functions $r$ and $s$. After a lengthy calculation we find that the commuting transverse dynamical system to (\ref{l1}) is of the form
\begin{eqnarray}
\dot{x}&=&-(4m+2)x[1+\sum_{r=0}^m A_{mr}yx^{2r}(x^2+y^2)^{m-r}],  \nonumber \\
\dot{y}&=&-(4m+2)y[1+\sum_{r=0}^m A_{mr}yx^{2r}(x^2+y^2)^{m-r}].\label{csma}
\end{eqnarray}

Since (\ref{csma}) also admits the same inverse integrating factor as that of its commuting system (\ref{l1}), one can determine the first integral of the former by evaluating the integral (\ref{eq2}). Doing so we obtained exactly the same form of the first integral as in the sub-cases, that is Eq.(\ref{nb1}). Finally, fixing $f(H)=H$ and $g(I)=\frac{y} {x}$, we find the linearizing transformation is of the form
\begin{eqnarray}
u&=&\frac{y[1+\sum_{r=0}^n A_{mr}yx^{2r}(x^2+y^2)^{m-r}]} {(x^2+y^2)^{m+1}},\nonumber \\
v&=&\frac{x[1+\sum_{r=0}^n A_{mr}yx^{2r}(x^2+y^2)^{m-r}]} {(x^2+y^2)^{m+1}}.\label{bel1}
\end{eqnarray}
We verified that under this transformation the original nonlinear equation (\ref{l1}) becomes a linear harmonic oscillator equation (\ref{har1}). Substituting the general solution of the latter in the left hand side in (\ref{bel1}) and rewriting the resultant expressions for $x$ and $y$ we arrive at the general solution of (\ref{v1}), that is
\begin{eqnarray}
x(t)&=&\frac{\sin(t-t_0)} {(C-\cos(t-t_0)\sum_{r=0}^nA_{mr}\sin^{2r}(t-t_0))^{1/(2m+1)}},\nonumber \\
y(t)&=&\frac{\cos(t-t_0)} {(C-\cos(t-t_0)\sum_{r=0}^nA_{mr}\sin^{2r}(t-t_0))^{1/(2m+1)}},
\end{eqnarray}
where $C$ and $t_0$ are the integration constants and $A_{mr}$ is given in Eq.(\ref{amr}).

\subsection{ Other linearizing transformations}
We have seen that the transformation (\ref{bel1}) transforms the system (\ref{l1}) into the harmonic oscillator equation. Equivalently, the nonlocal transformation $U=x e^{(\int x^{2m+1}dt)}$ also takes the equation (\ref{j1}) to the harmonic oscillator equation, $\ddot{U}+U=0$. Substituting the solution of the latter in the equation, $\frac{\dot{U}} {U}=\frac{\dot{x}} {x}+x^{2m+1}$ which is obtained by rewriting the nonlocal transformation in the above form, and solving the resultant equation for $x$ one ends up at the same solution as given in (\ref{bel1}). Interestingly Eq.(\ref{l1}) can also be transformed to the free particle equation $\frac{d^2F} {dG^2}=0$, through the generalized linearizing transformation (GLT)  $F=\phi{(t)}$, $G=\frac{\phi'{(t)}} {t+\tan^{-1}\big(\frac{\dot{x}} {x}+x^{2m+1}\big)}$, where $\phi$ is an arbitrary function of $t$. From the particular solution of the free particle equation one can derive the general solution of (\ref{j1}). The resultant expression again agrees with the one given in (\ref{bel1}).

\section{\bf Conclusion}
In this paper we have studied in detail the linearization of isochronous centres of a modified Emden equation with linear external forcing, namely (\ref{l1}). This is a continuation work of our earlier studies on exploring the classical dynamics of this model. The core idea we have employed here is to explore a transverse commuting dynamical system to the original equation and constructing first integral for both the systems. The necessary linearizing transformations can be deduced from these two integrals in a straightforward manner. Through this work we have brought out certain geometrical features associated with the system. To make our studies systematic to begin with we have considered case $m=0$ in (\ref{j1}) and then $m=1$ and finally consolidated the results for $m=arbitrary$. We have given the inverse integrating factor, first integral for the dynamical system under consideration and its transverse commuting system. In the course of this analysis we have also observed that the first integral associated with the transverse vector field is same irrespective of the value of $m$. We have also found the explicit form of linearizing transformation for the Eq.(\ref{l1}). For the sake of completeness we have also given the general solution.

\section*{Acknowledgements}
RMS acknowledges the University Grants Commission (UGC-RFSMS), Government of India, for providing a Research Fellowship and the work of MS forms part of a research project sponsored by Department of Science and Technology, Government of India.

\end{document}